# Epitaxial synthesis of unintentionally doped *p*-type SnO (001) via *suboxide* molecular beam epitaxy

Kingsley Egbo,[1,*] Esperanza Luna,[1] Jonas Lähnemann,[1] Georg Hoffmann,[1] Achim Trampert,[1] Jona Grümbel,[2] Elias Kluth,[2] Martin Feneberg,[2] Rüdiger Goldhahn,[2] Oliver Bierwagen[1,†]

[1] *Paul-Drude-Institut für Festkörperelektronik, Leibniz-Institut im Forschungsverbund Berlin e.V., Hausvogteiplatz 5–7, 10117 Berlin, Germany*

[2] *Institut für Experimentelle Physik, Otto-von-Guericke-Universität Magdeburg, Universitätsplatz 2, 39106 Magdeburg, Germany*

*egbo@pdi-berlin.de

[†]bierwagen@pdi-berlin.de

**ABSTRACT**

By employing a mixed $SnO_2$+Sn source, we demonstrate *suboxide* molecular beam epitaxy growth of phase-pure single crystalline metastable SnO(001) thin films on YSZ (001) substrates at a growth rate of ~1.0 nm/min without the need for additional oxygen. These films grow epitaxially across a wide substrate temperature range from 150 to 450 °C. Hence, we present an alternative pathway to overcome limitations of high Sn or $SnO_2$ cell temperatures and narrow growth windows encountered in previous MBE growth of metastable SnO. In-situ laser reflectometry and line-of-sight quadrupole mass spectrometry were used to investigate the rate of SnO desorption as a function of substrate temperature. While SnO ad-molecules desorption at $T_S$ = 450°C was growth-rate limiting, the SnO films did not desorb at this temperature after growth in vacuum. The SnO(001) thin films are transparent and unintentionally *p*-type doped, with hole concentrations and mobilities in the range of 0.9 to $6·10^{18}$ cm$^{-3}$ and 2.0 to 5.5 cm$^2$V$^{-1}$s$^{-1}$, respectively. These p-type SnO films obtained at low substrate temperatures are promising for back-end-of-line (BEOL) compatible applications and for integration with *n*-type oxides in *p-n* heterojunctions and field-effect transistors.

# I. INTRODUCTION

Tin (II) oxide (SnO) is a valuable *p*-type oxide material useful in several technological applications such as in *p-n* diodes, transistors, solar cells and solid-state gas sensing[1,2]. It is among the few binary oxide materials that show unintentional *p*-type character due to native defects[3–7]. Though, SnO has a low indirect fundamental bandgap of 0.7 eV, an optical bandgap of ~2.7-2.9 eV makes it a suitable component for several transparent applications[8,9]. SnO crystallizes in a layered tetragonal litharge structure with space group P4/nmm, with four oxygen atoms and an Sn atom forming a pyramid structure[10,11]. Due to a more dispersed valence band maximum composed of hybridized Sn 5*s* and the O 2*p* orbitals[12,13] reasonably high mobilities compared to most *p*-type oxides have been reported for polycrystalline and single crystalline SnO thin films[14–16]. As most metal oxides show a propensity for *n*-type conductivity and due to the difficulty in their bipolar doping, efforts in oxide electronics mostly depends on oxide heterojunctions. Hence, SnO has become a major *p*-type oxide for oxide heterojunctions due to its high mobility compared to other *p*-type oxides[17–21].

Unlike Tin (IV) oxide ($SnO_2$), SnO is metastable, this presents a growth challenge with competing stable metallic Sn or $SnO_2$ as well as $Sn_3O_4$ phases which may coexist along with SnO phase in the thin films. Though, several growth techniques have been employed for the growth of polycrystalline SnO,[20] phase-pure single crystal SnO layers have been mostly obtained by electron beam evaporation[22], pulsed laser deposition[8] and molecular beam epitaxy. High quality single crystalline SnO has been grown by plasma assisted MBE (PA-MBE) using a metal Sn source[23] and S-MBE using an oxide $SnO_2$ source[24]. However, both methods present challenges; the plasma assisted growth of SnO using a Sn metal source requires the Sn effusion cell to be operated at very high temperatures up to 1175°C and the growth window is limited by the formation of $SnO_2$ and oxygen rich Sn compounds such as $Sn_3O_4$, requiring a rigorous fine tuning of Sn/O flux ratio.[23] Also, growth using an oxide $SnO_2$ charge involves high temperature decomposition of the oxide into the SnO suboxide, leading to a parasitic high oxygen background. Previously, it has been proposed that the sublimation of a mixed oxide + metal charges in an effusion cell can provide an effective suboxide flux for the growth of oxides by MBE [25]. For instance, we showed that a reaction of $SnO_2$ with Sn metal charge to give SnO ($SnO_2$ + Sn → 2SnO) can be an efficient source of SnO flux, for suboxide MBE growth of $SnO_2$ [25]. This suboxide approach also offers the advantage that the reactions; $SnO_2$ + Sn → 2 SnO takes place at lower effusion-cell temperatures and growth of SnO can proceed without the need for plasma activated oxygen or any intentional background oxygen.

Following this *suboxide*-MBE (S-MBE) approach, we present in this study the growth of SnO thin films without the need for plasma activated oxygen or intentional background oxygen. SnO is grown heteroepitaxially on Y-stabilized $ZrO_2$; [YSZ (001)] and r-plane $Al_2O_3$ in a substrate temperature, $T_S$ window from 50°C to 650°C. We investigate the impact of $T_S$ on the phase purity, growth-rate-limiting SnO desorption as well as structural and transport properties of the SnO layers. A schematic comparison of the conventional method of PA-MBE using a metal source and the S-MBE approach employed in this work is described in Figure 1(a) and (b) respectively. Detailed investigation of the growth kinetics for our SnO growth using this S-MBE approach indicate that the growth rate is limited by the desorption of the adsorbed SnO molecules which increases with the substrate temperature, however, grown single SnO crystalline layers are stable and show negligible desorption at substrate temperatures ≤450°C. We show that while amorphous layers were obtained at $T_S$ =50°C, textured and single-crystalline phase-pure unintentionally (UID) doped p-type SnO(001) films are grown between substrate temperatures of 150°C and 450°C. Above 550°C, secondary $Sn_3O_4$ and Sn phases are present in the SnO layer. Obtained single crystalline SnO(001) layers at low $T_S$ between 150-250°C show good UID transport properties. It is well known, that epitaxial and single crystalline channel layers result in superior qualities in devices such as thin film transistors, however, epitaxial deposition of active layers usually require high substrate temperatures that exceed the BEOL limit. Hence, these epitaxially grown SnO thin films obtained for $T_S$ between 150-250°C can be promising for applications as active layers for BEOL compatible device development.

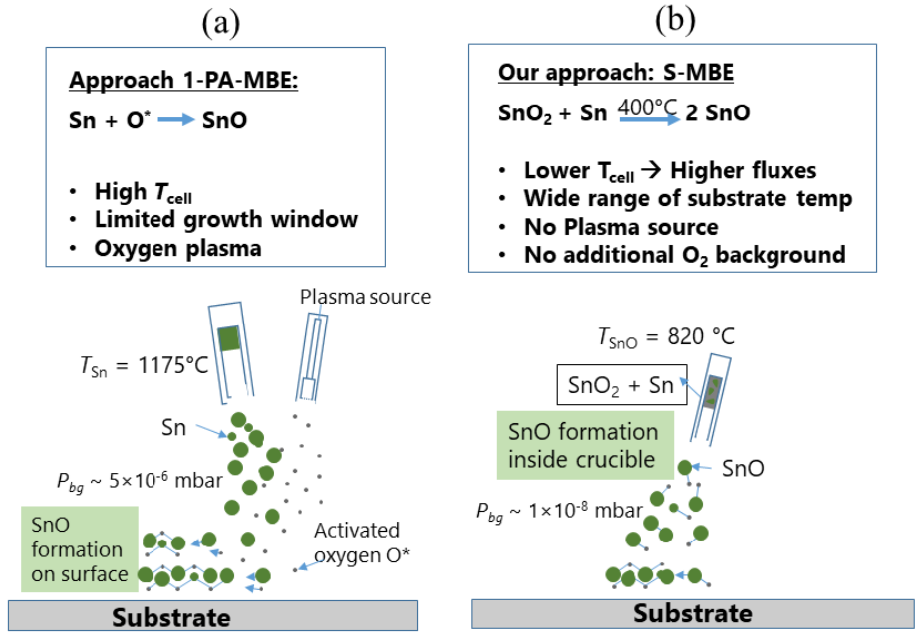

Figure 1: (a) Schematic of the formation of SnO layers on a substrate using the PA-MBE approach with Sn charge and (b) S-MBE approach using mixed $SnO_2$+Sn charge. In the PA-MBE approach, a Sn metal cell operated at very high cell temperature supplies Sn metal fluxes and a plasma source provides activated oxygen leading to the formation of SnO on the film surface, while in the S-MBE approach SnO is formed in the crucible by the mixture and films are deposited in high vacuum without activated or molecular oxygen present.

## II. EXPERIMENTAL

Approximately 100-190 nm-thick UID SnO(001) thin films were deposited in an MBE chamber with solid-source effusion cells. While most growth were performed on YSZ(001), r-plane $Al_2O_3$ substrates were also co-loaded for several growth runs. To grow SnO, a $SnO_2$+Sn mixed source was sublimed from an effusion cell with an $Al_2O_3$ crucible at temperatures between 740-820°C. The hot-lip of the used dual-filament cell was kept at 150°C above the $SnO_2$+Sn–cell temperature. The resulting source beam equivalent pressure (BEP), proportional to the particle flux and measured using a nude filament ion gauge positioned at the substrate location, is about 0.4-1.5 $\times 10^{-7}$ mbar for all growth runs. Before growth, the substrates, with 1 μm thick Ti sputter-deposited on the backside to improve substrate radiative heating, were plasma treated for 30 min at substrate temperatures, $T_S$ between 400-700°C using 1 SCCM $O_2$ and 200 W plasma power in the growth chamber. The $T_S$ monitored *in-situ* by a thermocouple between the substrate and heating filament was varied between 50-650 °C for different growth runs. The background pressure of the chamber during SnO deposition was maintained at $P_{GC}$ ~5-8 $\times 10^{-8}$ mbar without plasma activated oxygen and any intentional molecular oxygen. The growth rate and the amount of desorbing flux were measured *in-situ* by laser reflectometry (LR) and line-of-sight quadrupole mass spectrometry (QMS) [26].

Different *ex-situ* techniques were used to characterize the grown SnO layers. A 4-circle x-ray lab-diffractometer (X'pert Pro MRD from Philips PANalytical) equipped with a Cu Kα radiation source was used to investigate the crystallographic orientation of the film and the epitaxial relationship with the substrate. The out-of-plane orientation was analyzed by means of symmetric on-axis 2Θ-ω scans with a 1mm detector slit. The in-plane epitaxial relationship between the SnO film and the YSZ substrates were measured by $\Phi$-scans in a skew symmetric geometry. On-axis rocking curve ω-scans were used to investigate the crystalline quality of the films and texture maps in a skew-symmetric geometry were used to validate the phase purity. Bulk-sensitive room temperature Raman spectroscopy measurements in the backscattering geometry using a solid state laser at a wavelength of 473 nm were used to investigate the grown layers as described in Ref. [23]. Surface morphology of the grown layers was analyzed by atomic force microscopy (AFM) using a Bruker Dimension Edge in the peak force tapping mode. (Scanning) Transmission Electron Microscopy, (S)TEM was used to study the film's microstructure. TEM observations were made with a JEOL 2100F microscope operating at 200 kV. Cross-sectional TEM specimens were prepared for observation in both ⟨011⟩ projections of the YSZ substrate using standard mechanical polishing and dimpling, followed by argon ion-milling, starting at 3.0 keV and finishing at 1.5 keV. The film domain structures, and orientation was further investigated by electron backscatter diffraction (EBSD) measurements in a scanning electron microscope operated at 15 kV. Hall measurements in the van der Pauw geometry were used to investigate the transport properties of the layers grown at different substrate temperatures. Because of the low mobilities obtained for some of the grown layers, a magnetic sweep method was used to extract reliable Hall coefficients[27]. To compare the optical properties of grown layers with existing literature, spectroscopic ellipsometry measurement and modelling was performed on a layer grown on r-plane $Al_2O_3$ at a substrate temperature of 400°C.

## III. RESULTS AND DISCUSSIONS
### A. Thermodynamics of the suboxide source and SnO growth window
Previously from a quadrupole mass spectrometry study, we have demonstrated that a reaction of a mixed oxide + metal charge can serve as an efficient source of suboxides such as $Ga_2O$ and SnO[25]. By taking advantage of this suboxide source, we determine the growth window and thermodynamic consideration for the growth of metastable *p*-type SnO from a $SnO_2$+Sn mixture[23,28]. Figure 2(a) shows the calculated SnO$_2$-Sn equilibrium phase diagram as function of stoichiometry, $n_{Sn}/(n_{Sn} + n_{SnO2})$ and temperature calculated at 10$^{-7}$ mbar typical for MBE growth without O$_2$ environment. This thermodynamic equilibrium diagram calculated using FactSage[29] indicate that stable SnO in the solid phase can be obtained at growth temperatures within 190-420°C at a stoichiometry of $n_{Sn}/(n_{Sn} + n_{SnO2})$=0.5. Here ideal gas indicate gaseous species of the constituent elements in the reaction[25]. This supports our experimental data where phase pure SnO was obtained between 150-450°C substrate temperatures during growth as discussed below. Due to the absence of reactive oxygen during growth, we can obtain phase-pure SnO for Sn stoichiometry of 0.5 in $n_{Sn}/(n_{Sn} + n_{SnO2})$ which was used throughout this growth experiment. Figure 2(b) shows the corresponding phase diagrams for the mixed charges at Sn stoichiometry, $n_{Sn}/(n_{Sn} + n_{SnO2}) = 0.5$ indicating the vapour pressure of suboxide SnO as a function of source temperatures for the mixed charges. These mixed sources promote the availability of required suboxide vapour pressure at lower source temperature compared to the metal charge and solid oxide charge[25].

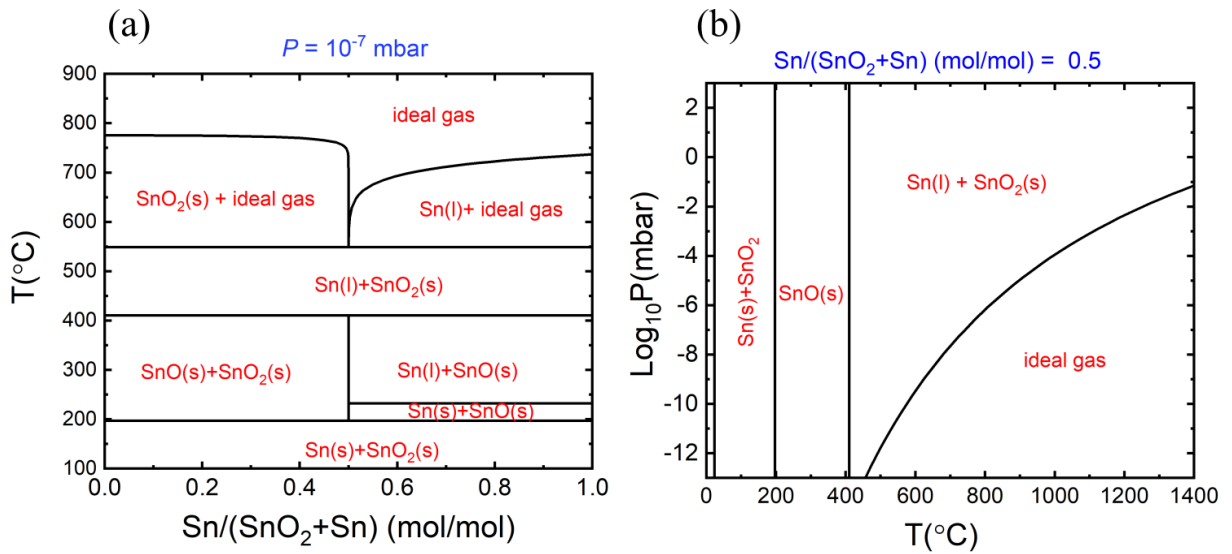

Figure 2: (a) Equilibrium phase diagram of the SnO$_2$-Sn system as a function of stoichiometry and temperature at a pressure of 10$^{-7}$mbar typical for MBE growth without intentional oxygen background. Stoichiometries of $n_{Sn}/(n_{Sn} + n_{SnO2}) = 0$, 0.5 and 1 correspond to SnO$_2$; SnO and pure Sn respectively. (b) Phase diagram of the SnO$_2$-Sn system for $n_{Sn}/(n_{Sn} + n_{SnO2})= 0.5$ as a function of temperature and pressure.

## B. SnO Flux, Rate of Desorption, and disproportionation of layer

In the identified growth window for SnO suboxide growth, the growth rate is given by the difference of the amount of arriving SnO species (proportional to the vapor pressure of the SnO [c.f. Fig. 2(b)] at the cell temperature of the mixed source) and the amount of desorbing species (which scales with the substrate temperature). This promotes a simpler growth kinetics compared to SnO growth by PA-MBE in which rigorous Sn/O flux calibration is required to limit the formation of competing O-rich and Sn-rich phases. The SnO flux impinging on a substrate from the mixed $SnO_2$+Sn effusion cell is equivalent to the measured BEP in the absence of an oxygen background. In order to characterize the kinetics of the mixed $SnO_2$+Sn effusion cell, we measured the BEP as a function of source temperature from 740-800°C in the absence of any active $O_2$ flow (see Fig. 3). The curve shows the expected exponential dependence, with an activation energy for SnO of 2.4 eV, similar to the value obtained by our previous QMS results[25]. The incorporated cation flux on the substrate (solid symbols in Fig. 3, obtained as the product of the measured growth rate and the cation density), is proportional to the BEP as expected for full cation incorporation due to the low growth temperature of 50°C, i.e., without desorption from the substrate.

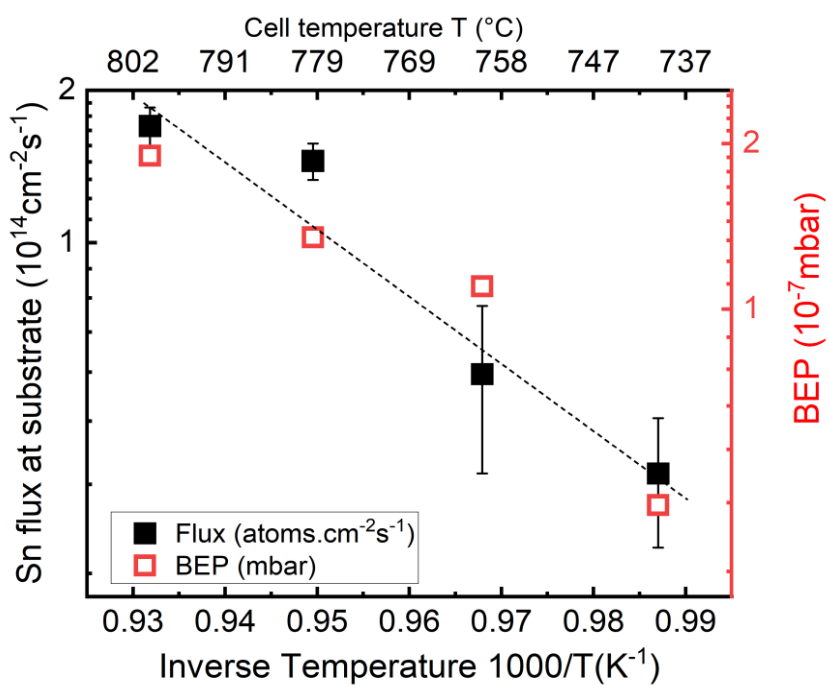

Figure 3: Arrhenius diagrams of the incorporated Sn flux (solid symbols) during MBE growth and the corresponding measured BEPs of the mixed $SnO_2$+Sn effusion cell (right axis, open symbols). An incorporated flux at the substrate of $4.8\times10^{13}$ $cm^{-2}s^{-1}$ corresponds to a SnO growth rate of 1.0 nm/min and to a BEP of ~$7\times 10^{-8}$ mbar.

Using line-of-sight quadrupole mass spectrometry, we identify the rate of desorption of the SnO ad-molecules during growth as function of substrate temperatures. Our measurement reveals that the desorbing flux significantly decreases with substrate temperature as shown in Figure 4(a).

The inset of Figure 4(a) shows typical QMS spectrum recorded during SnO desorption at high substrate temperature. For a flux of $4\times10^{13}$ cm$^{-2}$s$^{-1}$ reaching the substrate with substrate temperature of 450°C, a growth rate of 1.0 nm/min is expected, however an SnO growth rate of ~0.6nm/min is obtained from the LR oscillation in Figure S1, hence the desorption rate of the SnO ad-molecules is ~0.3 nm/min ($1.4\times10^{13}$ cm$^{-2}$s$^{-1}$). From the plot of desorption rate as a function of substrate temperature, an activation energy of desorption of ~0.3 eV is obtained, this low activation energy value indicates a high volatility of the suboxide ad-molecules during growth. The desorption rate at higher substrate temperatures is limited by the amount of flux reaching the substrate as seen in Figure 4(a). While these ad-molecules are volatile during growth, to further understand the stability of the SnO molecules within the film after growth, we perform a separate LR study at a SnO film grown at a BEP of $1.3\times10^{-7}$ mbar. From Figure 3, this BEP corresponds to a flux at the substrate of $8.6\times10^{13}$ cm$^{-2}$s$^{-1}$ (assuming full incorporation) and to a growth rate of 1.85 nm/min. However, Fig. 4(b) shows only a growth rate, of 1.2 nm/min obtained from the oscillatory half period $t_{growth}$ of the laser reflectometry signal[26] for growth at 450°C substrate temperature; which is the highest temperature where phase pure SnO is obtained. Hence, the SnO ad-molecules has a desorption rate of ~0.65 nm/min during growth corresponding to a desorbing flux of $3.0\times10^{13}$ cm$^{-2}$s$^{-1}$. When the SnO shutter is closed and the substrate temperature maintained at 450°C the reflected laser signal is constant indicating negligible desorption of stabilized SnO species on the substrate This is in contrast to the high desorbing flux of $3.0\times10^{13}$ cm$^{-2}$s$^{-1}$ for the SnO ad-molecules during growth at 450°C, indicating that once the SnO molecule stabilizes in a layer, its activation energy for desorption is higher than for directly re-evaporated SnO ad-molecules. Finally, with increase in substrate temperature to 550°C, an onset of an oscillation is observed due to SnO desorption or increasing roughness due to disproportionation of grown layer. Note that this reflected signal oscillation has weak amplitude compared to the growth oscillation pointing to the roughening of the layer and formation of other Sn-compound phases. Therefore while SnO ad-molecules desorb during growth as indicated in Figure 4(a), SnO layers formed are very stable with negligible desorption rate at growth temperature.

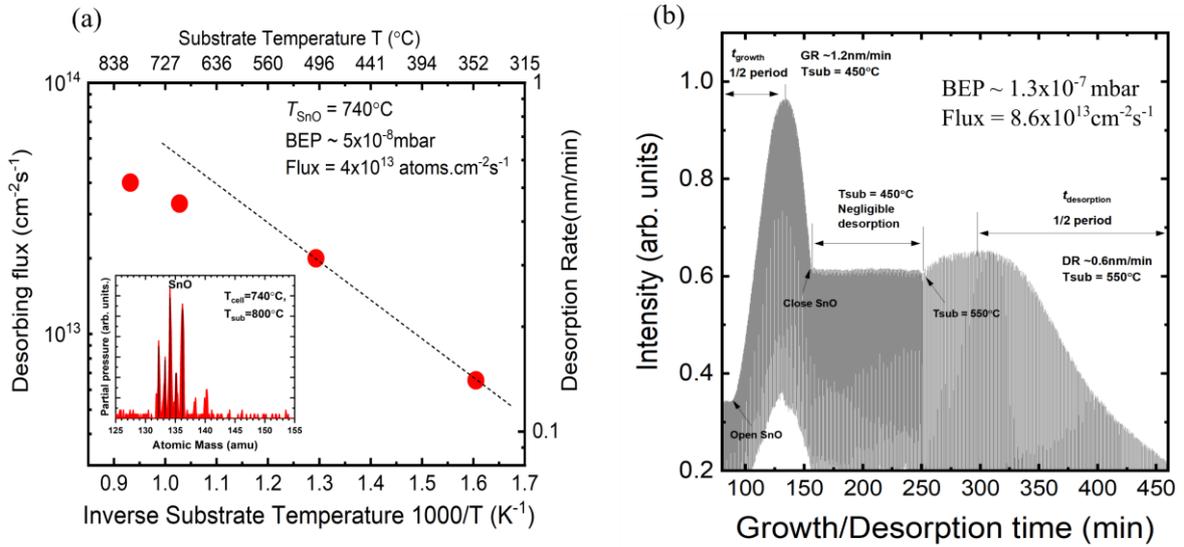

Figure 4. (a) Arrhenius diagrams of the desorbing flux as a function of the substrate temperatures, the desorbing species obtained from the line-of-sight QMS increases with increasing substrate temperatures and the desorbing flux is limited at high temperatures by the provided SnO flux to the substrates (b) Plot of the laser reflectometry signal (with substrate rotation) from the film surface during growth and thermal etching at higher substrate temperatures. The opening and closing of the $SnO_2$+Sn cell shutter is indicated.

## C. Structural Properties and Epitaxial Relation

Wide-angle symmetric 2Θ-ω XRD scans were used to investigate the out-of-plane orientation of the layers. Figure 5 (a) shows representative results for layers grown at different YSZ(001) substrate temperatures, wide-angle scans of selected samples between 10-120 deg are shown in Fig. S2 (a)(Supplementary Material), similar wide-angle scan for a sample grown on r-plane $Al_2O_3$ are shown in Figure S2(b). Typical streaky RHEED patterns observed during the growth of SnO (001) layers are also shown in Fig. S3 (Supplementary Materials). The layer grown at a substrate temperature of 50°C was amorphous, hence only substrate peaks are observed in the XRD scan. For layers grown between 150 to 450°C, only the SnO (001) and higher order reflexes as well as the YSZ (100) substrate peaks are present, indicating phase pure (001) oriented single crystalline SnO films. This is in contrast with previous results on plasma-assisted MBE growth of SnO in which phase-pure SnO was only possible at a substrate temperature between 350-400°C[23]. A very limited substrate temperature window was observed for previous S-MBE SnO growth on r-plane $Al_2O_3$ (1-102) substrates using $SnO_2$ source, Mei *et.al.,* reported that their films grown below 370°C were amorphous while no deposition occurred above 400°C[24]. Furthermore, we observe slightly sharper peaks in the 2Θ-ω scans of the samples grown at 250-450°C indicating higher crystal quality than that of layers grown at lower temperature. However, the increase is not linear with increasing substrate temperature. Once the substrate temperature is increased even further, to 550°C, the presence of $Sn_3O_4$ secondary phase and Sn peaks are observed indicating a disproportionation of the phase-pure SnO to Sn-rich and O-rich phases. This trend is continued with a further increase in substrate temperature up to 650°C where the intensity of a mixed phases (metallic Sn, $Sn_3O_4$) dominate the XRD and the SnO peak intensity becomes negligible. Figure 5(b) shows the FWHM of the SnO (002) omega rocking curves as a function of substrate temperature for the grown layers. The FWHM maximum ranges from a high value of 1.3° for the sample grown at a low substrate temperature of 150°C to values between 0.2-0.7 for samples grown at higher substrate temperatures. Film roughness between 2-15 nm (root mean squre, rms) are obtained from AFM measurements for the single crystalline SnO (001) layers.

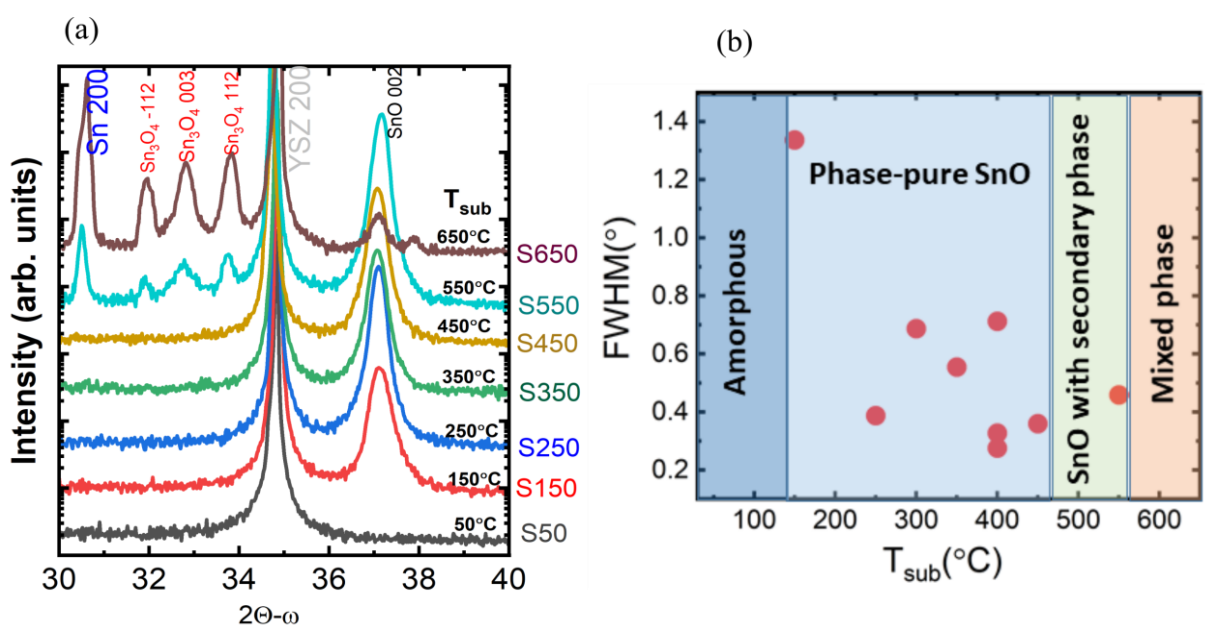

Figure 5(a) XRD out-of-plane symmetric 2Θ-ω scan of SnO(001) layer on YSZ(100) grown at different substrate temperatures. The film grown at 50°C was amorphous, between 150°C-450°C, phase pure single crystalline SnO(001) layers are obtained. At a substrate temperature of 550°C, secondary $Sn_3O_4$ phases are observed and growth at 650°C showed a mixed phase with negligible SnO(001) contribution. (b). FWHM of the SnO 002 omega rocking curve as a function of the substrate temperature.

The in-plane epitaxial relationship between the SnO(001) layer and the substrate was investigated for the phase pure SnO(001) thin films by skew-symmetric $Φ$-scans of the SnO 112 and YSZ 111 reflections with rotational angle $Φ$ around the surface normal (see supplementary figure S4). From Figs. 5(a) and S4, we confirm the epitaxial relationship SnO(001)∥YSZ(001) and SnO(110)∥YSZ(010) for the out-of-plane and in-plane directions, respectively, indicating therefore a 45° in-plane rotation of the SnO crystal lattice with respect to the YSZ one. This is in agreement with our previously grown PA-MBE thin films, where the same epitaxial relations are observed[23]. To further verify the in-plane crystalline texture of the grown layers, Fig. 6 shows the texture map of the SnO 101 reflex. The texture map confirms the presence of predominantly single domain as observed in the $\Phi$-scans. The scan shows four distinct circular peaks at a tilt angle of $\Psi = 51.63°$ corresponding to the tilt of the (101) plane with respect to the (001) planes. Other Streaky peaks not labelled observed in the texture map are due to the substrate holder as shown in Fig. S5 (Supplementary material). Texture maps also show that distinct in-plane epitaxial relationships are possible for these samples on YSZ(001) substrates unlike randomly oriented samples observed for growths on $Ga_2O_3$ substrates[17,18].

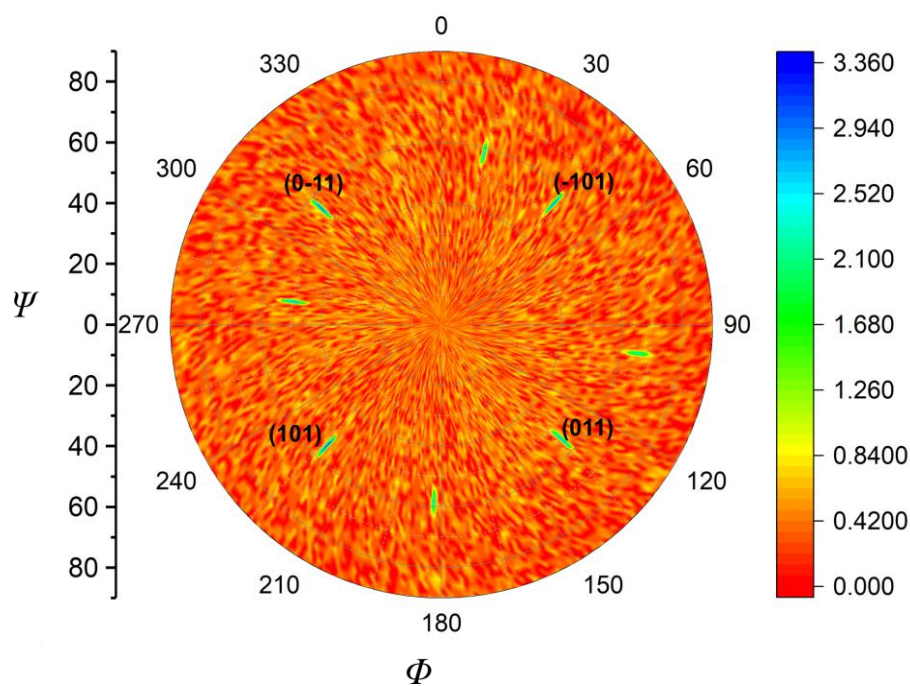

Figure 6 Texture scan along the (101) peak of SnO. Four peaks obtained for $\Phi$: 0-360° and $\Psi$: 0-90°. Peaks due to the (101) reflex are separated by 90° as expected for the Wulff plot in the supplementary material (Figure S7).

To clarify the phase composition of the grown epilayers beyond the limited capabilities of our out-of-plane XRD scans, room temperature bulk-sensitive Raman spectroscopy measurements are conducted on the grown layers. Following the peak assignment of Eifert *et. al.,*(and references therein) and our previously reported study on PA-MBE of SnO, the measured Raman spectra of all S-MBE grown SnO films were compared to phonon modes due to Sn, SnO and $Sn_3O_4$[23,30,31] as shown in Fig. 7. Raman spectra of the single crystalline samples grown between 150-450°C show only SnO $B_{1g}$ (113 cm$^{-1}$) and $A_{1g}$ (211 cm$^{-1}$) peaks strongly suggesting that no secondary phases are present. In agreement with the x-ray diffraction data, bulk-sensitive Raman scattering of the sample grown at 550°C indicates a weak contribution of the $Sn_3O_4$ phase coexisting with the SnO phase, while the sample grown at 650°C showed predominantly $Sn_3O_4$ peaks with weak contributions from Sn and SnO. These samples at 550°C also show slight shift in their SnO peak positions likely due to the presence of these secondary phases.

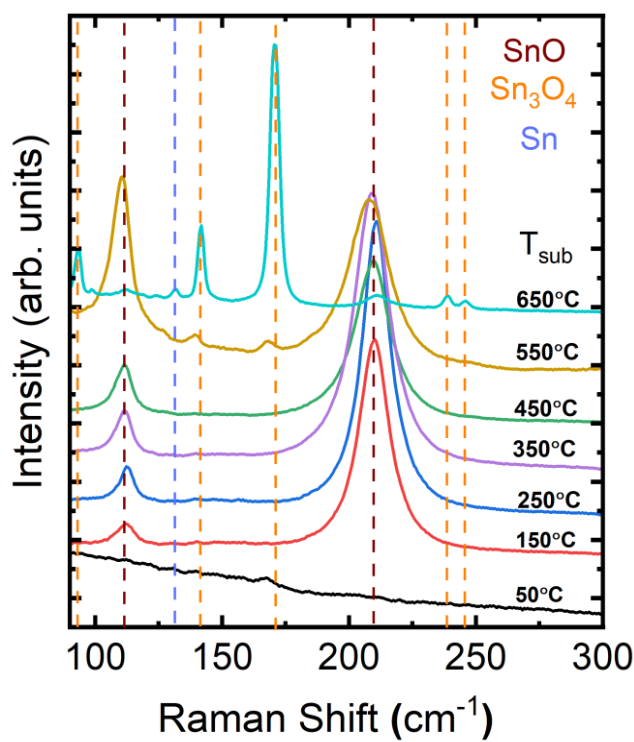

Figure 7. Bulk sensitive Raman spectra measured with an excitation wavelength of 473 nm. Vertical lines indicate the peak positions of dominant Raman active phonon modes, in SnO $B_{1g}$ (113) and $A_{1g}$ (211) and in $Sn_3O_4$, $A_g$ and $B_g$ as indicated by the colour code.

Transmission electron microscopy (TEM) of the sample cross-section is used to investigate the microstructure of the SnO layer grown at 400°C, its interface to the YSZ substrate and the epitaxial relationship between the YSZ substrates and the SnO layers. The overview bright field image of the SnO films as shown in Figure 8(a) reveals a layer composed of coalesced grains (their average diameter is about 150 nm), a morphology that suggests a 3-D Volmer-Weber growth mode of SnO on YSZ. The streak RHEED patterns obtained during growth shown Figure S3 is hence likely due to reflections from the surface of these large grains and does not imply a layer-by-layer growth mode. The grains have a single orientation and a well-defined epitaxial relationship. Figure 8(b) displays a high-angle annular dark-field (HAADF) STEM image of the SnO(001)/YSZ(001) heterostructure with atomic number Z-contrast. In this image, SnO shows a brighter contrast compared to YSZ due to its higher average Z. Figure 8(c) display high resolution TEM (HRTEM) phase-contrast micrographs of the SnO/YSZ(001) interface acquired along the [011] zone axis of the substrate. Though HRTEM reveals the local presence of steps at the SnO/YSZ interface, there is no noticeable misalignment of the SnO(001) layer which grows epitaxially on YSZ(001) following the epitaxial relationship $(001)_{SnO} \parallel (001)_{YSZ}$ and $[010]_{SnO} \parallel [1\text{-}10]_{YSZ}$. The perfect epitaxial alignment is reflected in the Fast Fourier Transform (FFT) pattern of the image in 8(d).

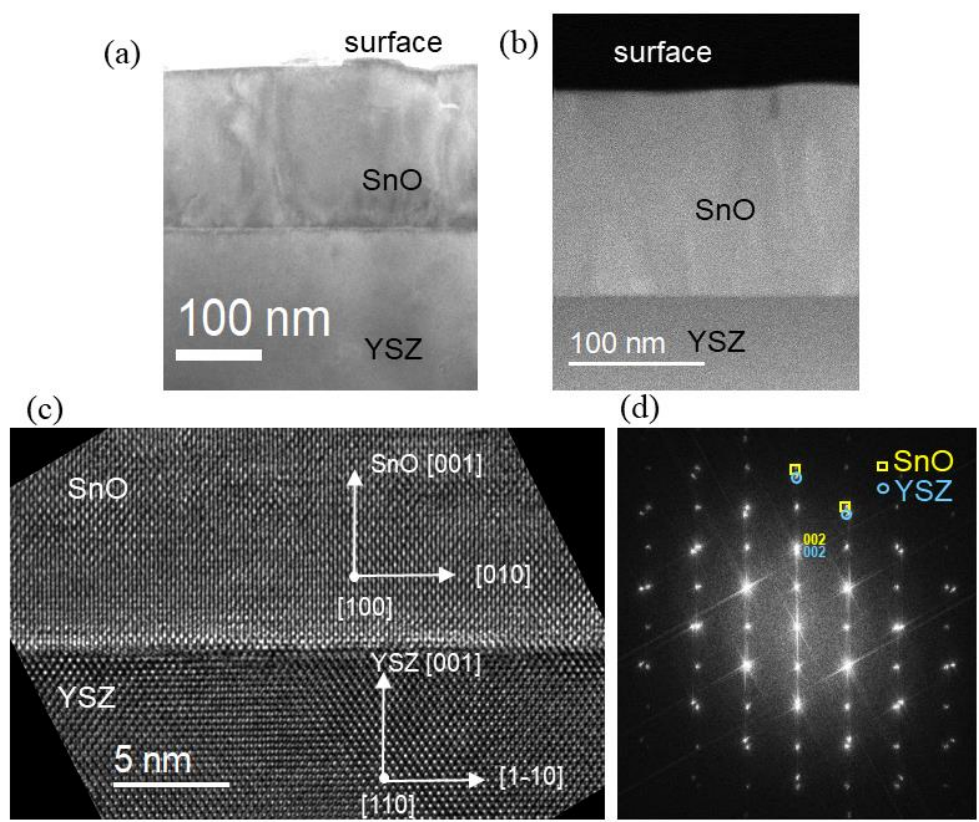

Figure 8: (a) Cross-sectional TEM image of the SnO(001) film grown at 450°C (b) High-angle annular dark-field (HAADF) STEM Z-contrast image of the SnO(001)/YSZ(001) heterostructure. (c), (d) High-resolution transmission electron microscopy (HRTEM) images of the SnO(001) epilayer on YSZ(001) near the SnO/YSZ interface acquired along the [011] zone axis of YSZ(100). Indexed reflections indicate an $(001)_{SnO} \parallel (001)_{YSZ}$ and $[010]_{SnO} \parallel [1\text{-}10]_{YSZ}$ epitaxial relationship as shown in (e) the FFT of the SnO/YSZ micrograph

Electron backscattering diffraction (EBSD) measured plan-view in a scanning electron microscope was further used to investigate the surface microstructure of our grown layers as the probing depth of EBSD is limited to ~20 nm. Figure S6 (a) show EBSD mapping of the SnO(001) indicating uniformity of the crystal orientation without any crystal twins on the layer. Figure S7 (b) show Kikuchi patterns from EBSD measurement and wireframe representation of the indexed orientations.

**D. Electrical Properties of S-MBE grown UID SnO(001) thin films.**

The charge carrier transport properties of the UID SnO (001) thin films were obtained at room temperature by Hall measurements in the *van der Pauw* geometry. For the van der Pauw measurement, indium contacts are made on the corners of the as grown UID SnO thin films prior to the measurement[32]. Due to low hole mobilities observed in most p-type oxides, extracting accurate Hall voltage using a single magnetic field value $B$ may become ambiguous. Here, to extract reliable Hall coefficients, a Hall sweep between +0.8T to -0.8T were carried out and the Hall coefficient is obtained from the slope of the Hall voltage as shown in Figure S7 (Supplementary Materials). Varying room temperature UID hole densities, $p_{Hall}$ in the range of 0.9-6.5x10$^{18}$ cm$^{-3}$ are obtained for the phase-pure SnO thin films (S150-S450) as shown in Figure 9(a). The sample grown at low substrate temperature of 150°C shows a remarkable mobility of 2.2 cm$^2$/V.s, increasing the substrate temperatures resulted in the growth of samples with average mobility of 4.5 cm$^2$/V.s as seen in Figure 9(b). Room temperature electrical resistivities $\rho$ between 0.3-1.2 Ohm-cm are obtained for the deposited films. A slight spread in transport properties for single crystalline SnO (001) grown at different temperatures maybe related to the growth temperature and crystallinity of the layers, but a consistent trend is not observed. Obtained hole densities in our grown layers are about 2 orders of magnitude higher than the value reported for UID SnO grown by S-MBE from an SnO$_2$ source[24]. Table 1 summarizes the room temperature electrical properties for various reported single crystalline p-type SnO(001) thin films grown using different techniques. The hole densities from these S-MBE grown layers are also slightly lower than our previously reported values grown via PA-MBE. This is likely due to the enhanced formation of Sn vacancies due to the energetics of the different growth process. For S-MBE growth, the SnO flux is reaching the substrate, this is expected to decrease the formation of Sn vacancies and complexes compared to the PA-MBE growth, where elemental Sn and activated oxygen is supplied. Nevertheless, we cannot rule out unintentionally incorporated extrinsic dopants. S550 films with dominant SnO also showed *p*-type UID properties like the phase pure SnO layers. Amorphous layers grown at 50°C was semi-insulating while the mixed phase layer grown at 650°C showed *n*-type character.

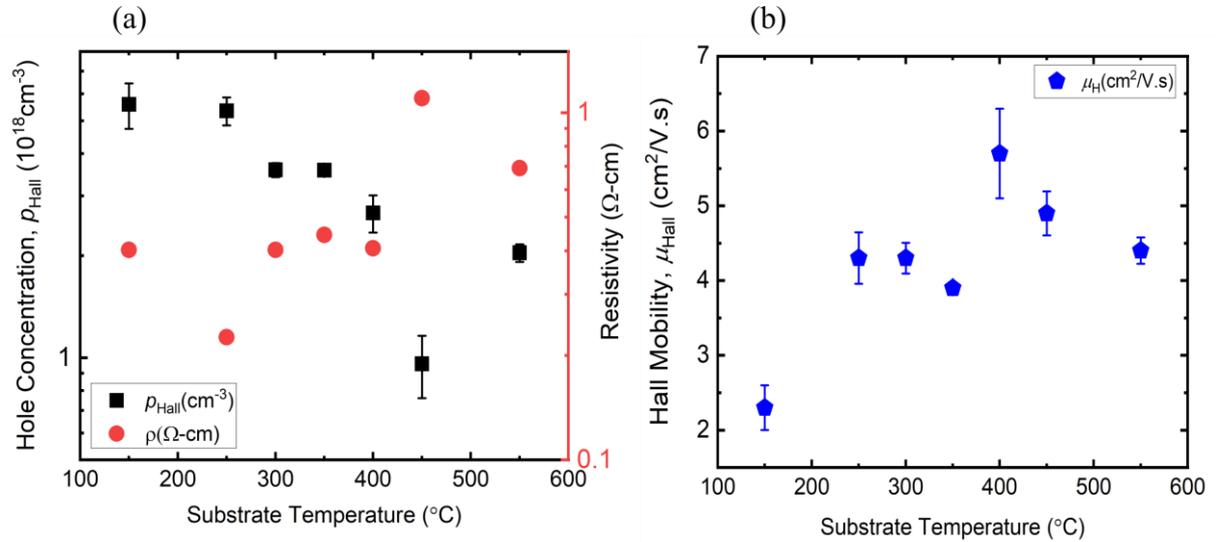

Figure 9(a) Hole concentration, *p* and resistivity, of phase-pure SnO (001) as a function of substrate temperature obtained from Hall measurements in the *van der Pauw* geometry, (b) Hall mobilities of deposited thin films as a function of substrate temperature.

Table 1: Comparison of electrical properties of epitaxial SnO(001) thin films grown using different techniques

| Material | Method[a] | $T_{dep}$ [°C] | Growth $P_{O2}$(Torr) | Substrate | FWHM(°) | $p_{Hall}$ [cm$^{-3}$] | Mobility(cm$^2$/V.s) | Resistivity(Ω-cm) | Ref |
|---|---|---|---|---|---|---|---|---|---|
| SnO(001) | S-MBE | 380 | 5x10$^{-7}$ | r-Al2O3 | 0.007 | 2.5x10$^{16}$ | 2.4 | 101 | 24 |
| SnO(001) | PLD | 575 | 1x10$^{-6}$ | YSZ(001) | 0.46 | 2.5x10$^{17}$ | 2.4 | - | 8 |
| SnO(001) | PLD | 200 | 6x10$^{-2}$ | YSZ(001) | 1.0 | 1.0x10$^{17}$ | 2.3 | - | 33 |
| SnO(001) | EBE | 600 | - | r-Al2O3 | 2.9 | 5.6x10$^{17}$ | 0.1 | 110 | 9 |
| SnO(001) | EBE | 600 | ~1x10$^{-6}$ | r-Al2O3 | | - | - | 195 | 22 |
| SnO(001) | PA-MBE | 350-400 | ~5x10$^{-6}$ | YSZ(001),c-Al2O3 | 0.4-1.9 | 1.8-9.7x10$^{18}$ | 1-6 | 0.25-2.0 | 23 |
| SnO(001) | S-MBE | 150-450 | ~6x10$^{-8}$ | YSZ(100) | 0.2-1.3 | 0.9-6 x 10$^{18}$ | 2.5-5.5 | 0.3-1.2 | This Work |
| SnO(001) | S-MBE | 400 | ~6x10$^{-8}$ | r-Al2O3 | 1.3 | 7.0x10$^{17}$ | 1.4 | 7.2 | This Work |

a) Method of deposition: S-MBE-Suboxide Molecular Beam Epitaxy, PA-MBE- Plasma assisted MBE PLD -Pulsed Laser Deposition, RFMS-RF Magnetron Sputtering, DCRS, DC Reactive Sputtering, EBE- Electron Beam evaporation

To assess the optical properties of SnO layers grown via this suboxide MBE route, spectroscopic ellipsometry measurement and modeling is performed on ~120 nm thick phase-pure SnO(001) sample grown on r-plane $Al_2O_3$ at 400°C. Fig S2(b) shows the wide angle 2Θ-ω scan. Compared to the sample grown on YSZ (001) at similar substrate temperature, Hall measurement for this sample show a slightly lower hole density and mobility of ~7.0x10$^{17}$cm$^{-3}$ and ~1.4 cm$^2$/V.s respectively and higher resistivity of ~7.2 Ohm-cm. This decrease in the transport properties is likely due to increase in strain-induced dislocation density caused by higher lattice mismatch of ~12% in SnO/r-plane $Al_2O_3$ hetero-interface compared to SnO/YSZ hetero-interface with ~5% lattice mismatch[23,24]. Ordinary and extraordinary complex dielectric function ($\varepsilon \equiv \varepsilon_1 + i\varepsilon_2$) spectra of this thin film extracted from room-temperature spectroscopic ellipsometry measurement is shown in Figure 10. The $\varepsilon_2$ spectra in the ordinary direction shows an onset of absorption at ~2.7 eV similar to previously reported values[24].

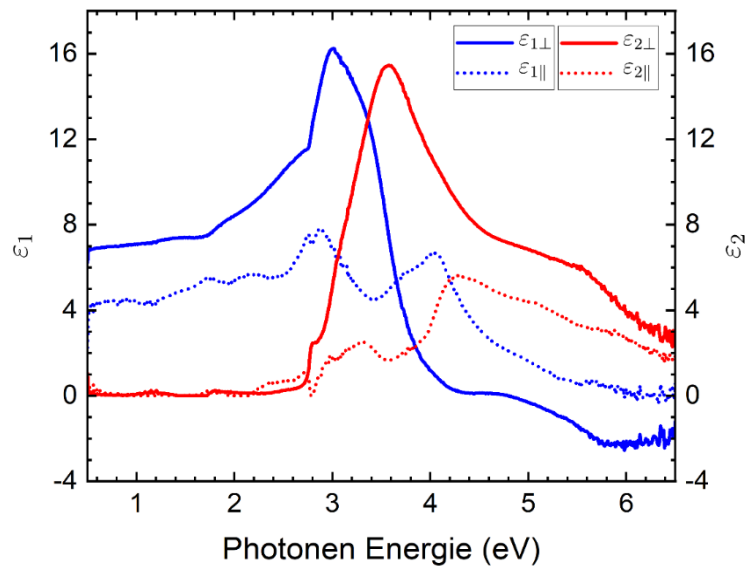

Figure 10: Complex dielectric function of SnO(001) thin film resolved into the ordinary *xy* (solid lines) and extraordinary *z* (dashed lines) components obtained from ellipsometry point-by-point fitting.

## IV. CONCLUSION

Using an intentional suboxide source comprising a mixed $SnO_2$+Sn charge, we demonstrate the heteroepitaxial growth of phase-pure, single crystalline SnO (001) thin films on YSZ(001) substrates. This S-MBE approach enabled the growth of phase-pure SnO(001) films across a wider growth window and substrate temperatures without plasma-activated oxygen or (un)intentional molecular oxygen. Hence overcoming the limitation of narrow growth window previously reported for PA-MBE growth of SnO using a metal charge and S-MBE growth of SnO using a $SnO_2$ source. We systematically characterized the S-MBE growth kinetics, such as the growth rate and desorbing SnO fluxes as a function of cell temperature, being an important step to employ this suboxide approach to other material systems. *Ex-situ* XRD and Raman measurements showed that phase-pure single crystalline SnO is obtained for a wide substrate temperature window between 150°C-450°C. Transport and optical measurement also confirm the p-type properties and optical transparency of these layers. Hence, with the S-MBE approach, single crystalline phase-pure SnO was achieved at the lowest substrate temperature of 150°C so far. This possibility to achieve epitaxial single crystalline p-type SnO(001) thin films for our S-MBE grown samples at low growth temperatures between 150-250°C can promote the integration of these p-type SnO layers for BEOL compatible device application.

## SUPPLEMENTARY MATERIAL

See supplementary material for typical RHEED image of SnO layers acquired during growth. Wide angle 2Θ-ω scans of SnO(001) on YSZ(001) and r-$Al_2O_3$. Skew-symmetric Φ-scans of SnO layer and YSZ substrate, EBSD mapping and Hall magnetic field sweep data.

## ACKNOWLEDGEMENTS

We would like to thank H.-P. Schönherr for MBE support, D. Steffen for TEM sample preparation, and P. John for critically reading the manuscript. This work was performed in the framework of GraFOx, a Leibniz ScienceCampus partially funded by the Leibniz Association.

## AUTHOR DECLARATIONS

Conflict of Interest

    The authors have no conflicts to declare

Supplementary Material for:

# Epitaxial synthesis of unintentionally doped *p*-type SnO (001) via *suboxide* molecular beam epitaxy

Kingsley Egbo,[1,*] Esperanza Luna,[1] Jonas Lähnemann,[1] Georg Hoffmann,[1] Achim Trampert,[1] Jona Grümbel,[2] Elias Kluth,[2] Martin Feneberg,[2] Rüdiger Goldhahn,[2] Oliver Bierwagen[1,†]

[1] *Paul-Drude-Institut für Festkörperelektronik, Leibniz-Institut im Forschungsverbund Berlin e.V., Hausvogteiplatz 5–7, 10117 Berlin, Germany*

[2] *Institut für Experimentelle Physik, Otto-von-Guericke-Universität Magdeburg, Universitätsplatz 2, 39106 Magdeburg, Germany*

*egbo@pdi-berlin.de

†bierwagen@pdi-berlin.de

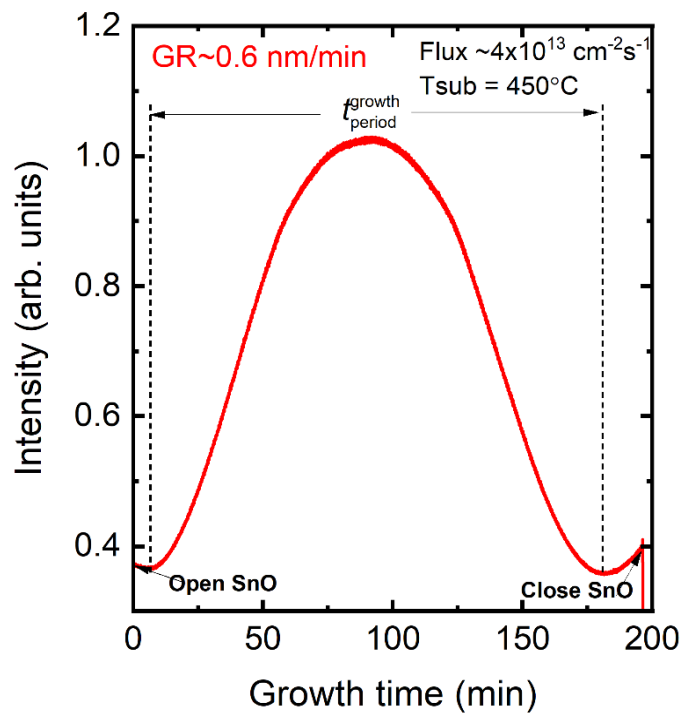

Figure S1: Laser reflectometry period signal for SnO thin film grown at a substrate temperature of 450°C and flux of $4 \times 10^{13}$ cm$^{-2}$s$^{-1}$. This corresponds to a growth rate of ~0.6nm/min. A 650 nm laser reflected at an angle of 60° with respect to the substrate normal was used for the laser reflectometry measurement.

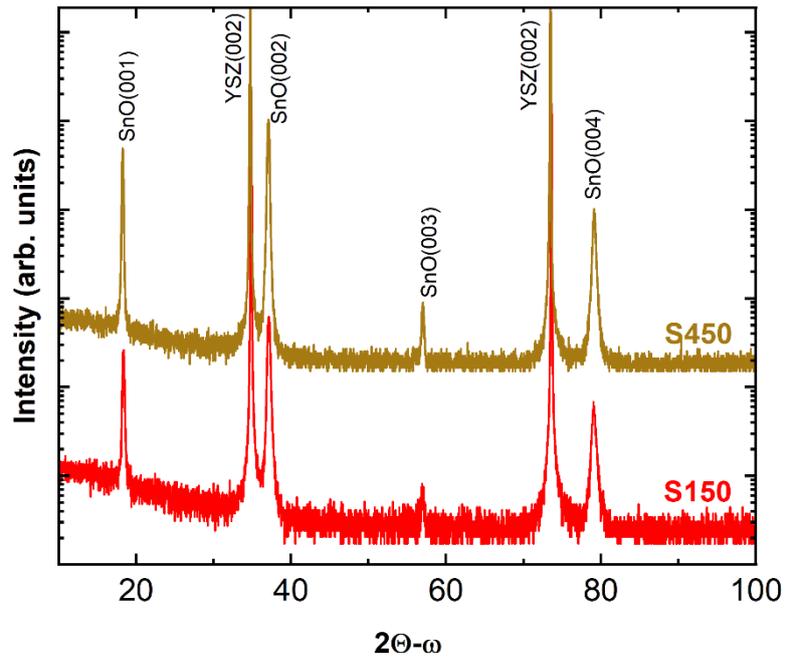

Figure S2(a): Wide angle 2Θ-ω scan of S150 and S450 SnO(001) thin films on YSZ(001).

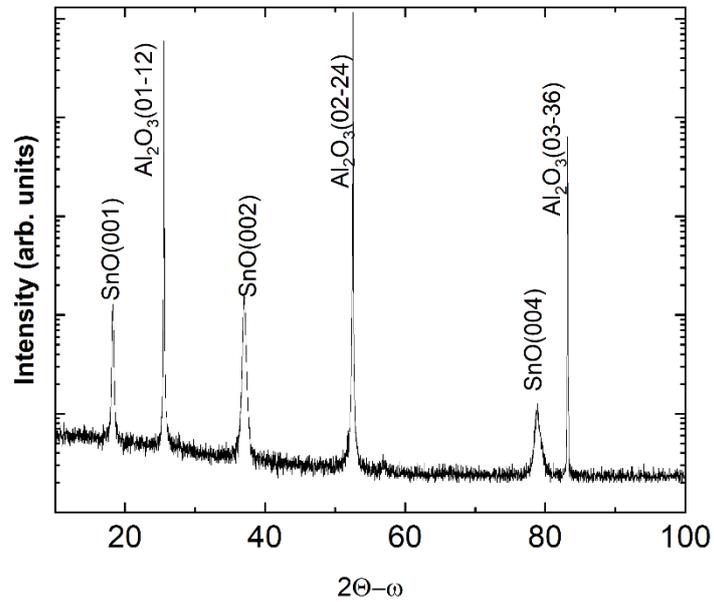

Figure S2(b): Wide angle 2Θ-ω scan SnO(001) layer grown on r-plane $Al_2O_3$ at 400°C

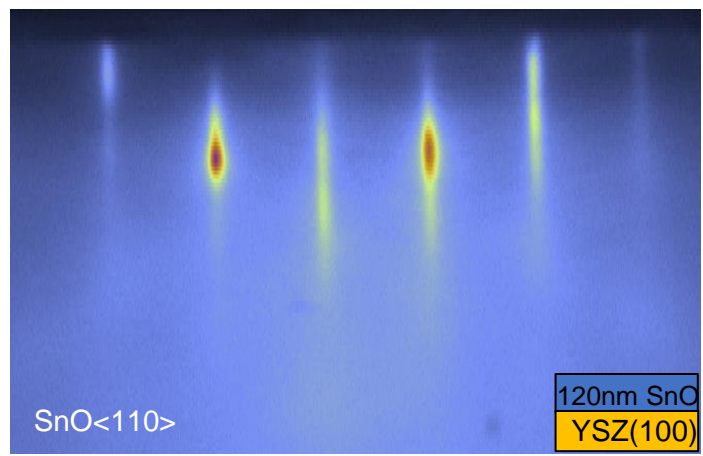

Figure S3: Typical streaky RHEED images suggest smooth surface with some islands during growth

The cubic four-fold rotational symmetry of the YSZ substrate is reflected by the four black peaks seen in Figure S4. The red peaks are the peaks due to the SnO(112) skew symmetric planes for S450 also showing 4 peaks due to the rotational symmetry hence indicating absence of multiple rotational domains in the epilayer[1]. The projection of the (111) peak of the YSZ onto the 100 plane is rotated by an angle of $\Phi = \pm 45°$ to the projection of the 112 peak unto the (001) plane of SnO.

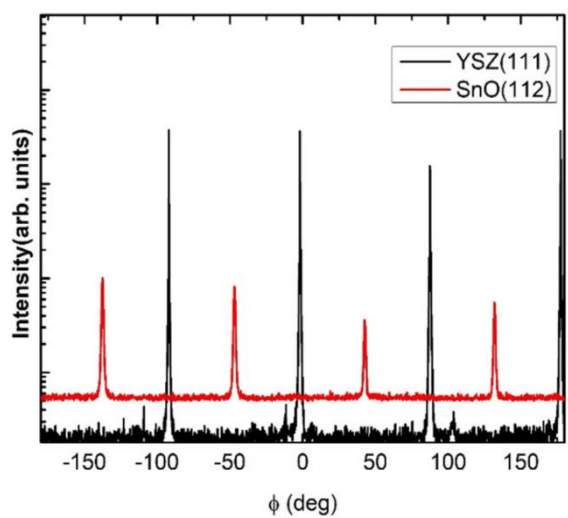

Figure S4: Phi-scans of the SnO and YSZ layers

SnO (001) Wulff plot and Texturemap of the substrate holder

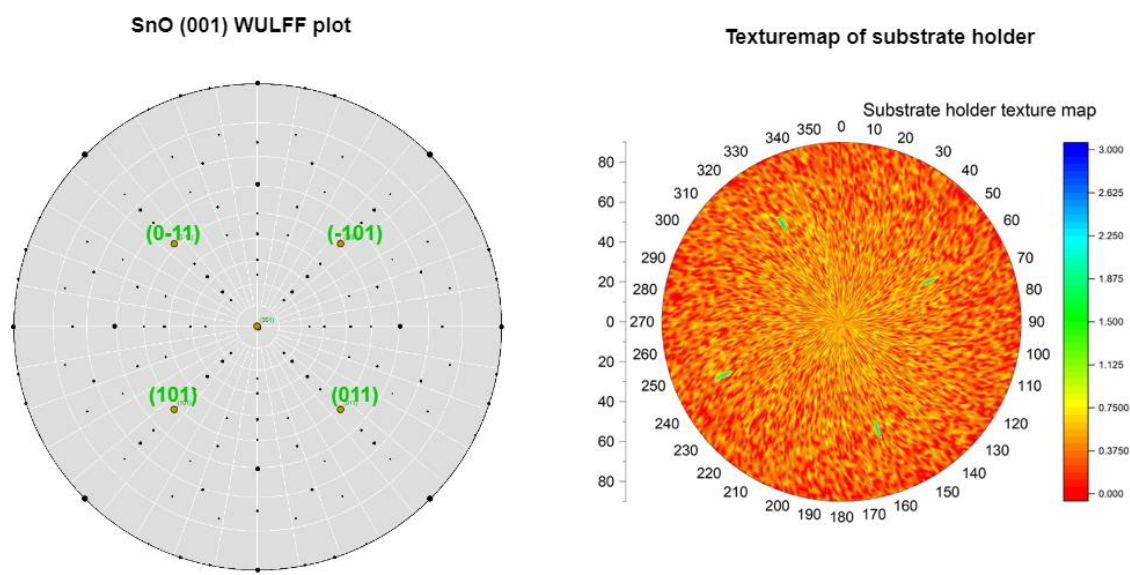

Figure S5: Calculated WULFF plot of the SnO(001) out-of-plane layer and Texturemap of substrate holder used in measurements.

EBSD Analysis of the crystal orientation in the surface of grown SnO layer and YSZ substrate

(a) 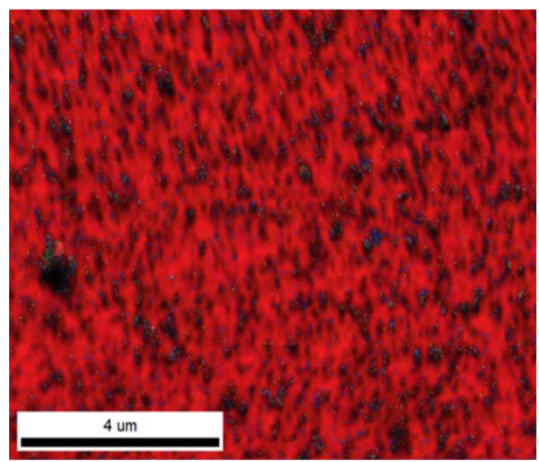  (b) 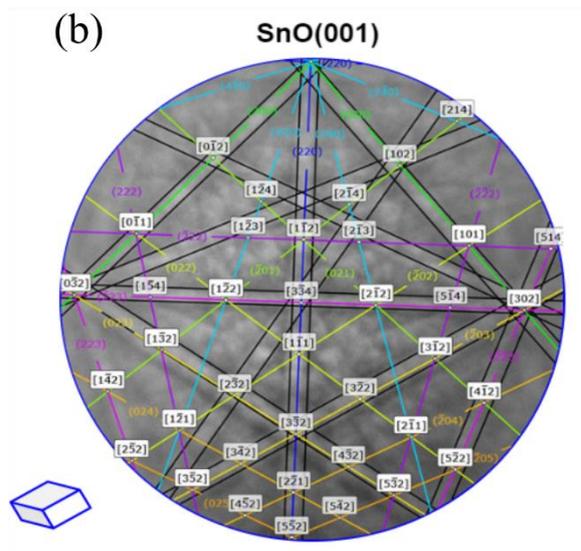

Figure S6. (a) EBSD Mapping of the SnO(001) layer showing uniformity across the surface of the sample (b) Kikuchi patterns of the SnO(001) epilayer and wireframe representation of the indexed orientations.

Hall Magnetic Field sweep

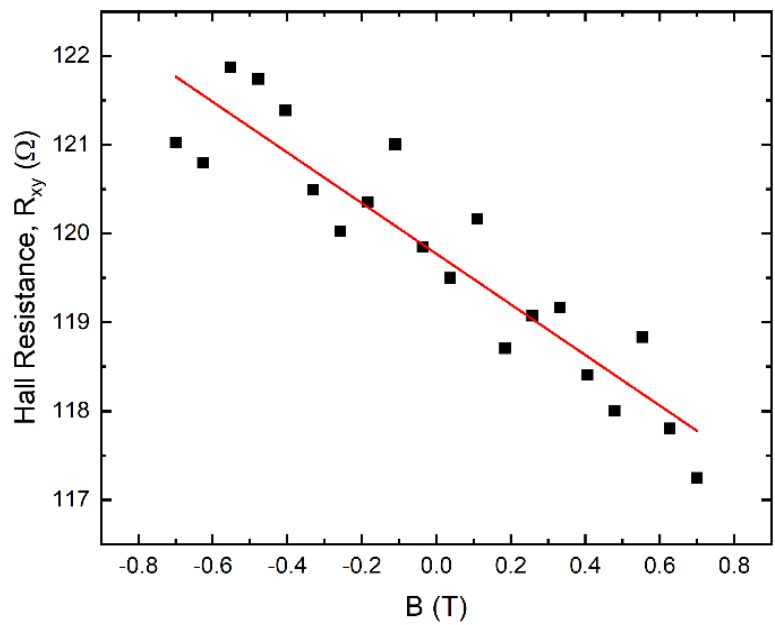

Figure S7. Typical magnetic field sweep used to extract the Hall coefficient of grown layers.

References

[1] M. Grundmann, T. Böntgen, and M. Lorenz, Phys. Rev. Lett. **105**, 146102 (2010).